\documentclass[superscriptaddress,twocolumn,prapplied]{revtex4}%
\usepackage{graphicx}
\usepackage{dcolumn}
\usepackage{bm}
\usepackage{color}
\usepackage{ulem}
\usepackage{amsmath}
\usepackage{amsfonts}
\usepackage{amssymb}
\usepackage{tikz}

\begin{document}

\title{Isotropic All-electric Spin analyzer based on a quantum ring with
spin-orbit couplings}

\author{Shenglin Peng}
\affiliation{State Key Laboratory of Powder Metallurgy, and Powder Metallurgy
Research Institute, Central South University, Changsha, P. R. China 410083}
\affiliation{School of Physics and Electronics, Central South University,
Changsha, P. R. China 410083}

\author{Wenchen Luo }
\email{luo.wenchen@csu.edu.cn}
\affiliation{School of Physics and Electronics, Central South University,
Changsha, P. R. China 410083}

\author{Jian Sun}
\affiliation{School of Physics and Electronics, Central South University,
Changsha, P. R. China 410083}

\author{Ai-Min Guo}
\affiliation{School of Physics and Electronics, Central South University,
Changsha, P. R. China 410083}

\author{Fangping Ouyang }
\email{ouyangfp@csu.edu.cn}
\affiliation{State Key Laboratory of Powder Metallurgy, and Powder Metallurgy
Research Institute, Central South University, Changsha, P. R. China 410083}
\affiliation{School of Physics and Electronics, Central South University,
Changsha, P. R. China 410083}
\affiliation{School of Physics and Technology, Xinjiang University, Urumqi,
P. R. China 830046}

\author{Tapash Chakraborty }
\email{Tapash.Chakraborty@umanitoba.ca}
\affiliation{Department of Physics and Astronomy, University of Manitoba,
Winnipeg, Canada R3T 2N2}

\date{\today }

\begin{abstract}
Here we propose an isotropic all electrical spin analyzer
in a quantum ring with spin-orbit coupling by analytically
and numerically modeling how the charge transmission rates
depend on the polarization of the incident spin. The
formalism of spin transmission and polarization rates in
an arbitrary direction is also developed by analyzing the
Aharonov-Bohm and the Aharonov-Casher effects. The topological
spin texture induced by the spin-orbit couplings essentially
contributes to the dynamic phase and plays an important role
in spin transport. The spin transport features derived
analytically has been confirmed numerically. This interesting
two-dimensional electron system can be designed as a spin
filter, spin polarizer and general analyzer by simply tuning
the spin-orbit couplings, which paves the way for realizing
the tunable and integrable spintronics device.
\end{abstract}

\maketitle

\section{Introduction}

Manipulation of the spin degrees of freedom and the conduction charges in
low-dimensional quantum structures has been attracting considerable interest,
due to wide range of potential applications in semiconductor spintronics and
quantum computation. How to control, modulate, or detect the spin
degree of freedom at the mesoscopic scale is a key step for the application
of the spin coherence in electronic devices. The quantum ring \cite{chapter,
ring1} is an ideal platform to take into consideration the Aharonov-Bohm
(AB) and the Aharonov-Casher (AC) effects to show the nature of the quantum
interference in conductance. The transport properties of similar nano-devices
have received considerable attention, especially in the spin transport device
subject to the Rashba spin-orbit coupling (SOC) \cite{datta,sun,chi,chang,
transport,chuang}, but the presence of Dresselhaus SOC or combination
of both SOCs\cite{miao,sil} have not been investigated sufficiently as yet.

The interplay of the Rashba SOC and the quantum interference has been widely
reported in the literature. No spin is being polarized \cite{Nitta,AC3,AC5}
in the transmission in the two-lead rings with equal arm length and without a
magnetic flux or an impurity. This is because in this case the interference
phase of the two different eigentransport channels is entirely due to the AC
effect. The signs of phases are opposite but the absolute values are equal
resulting in equal transmission rate for opposite spins. To polarize the
spin, we need to introduce magnetic fields \cite{Gumber,AC1,AC2,
Lucignano}, use unequal length arms \cite{Nitta,Wang,Tang}, doping
\cite{Bellucci,Citro,Kovalev}, or contact three or more leads \cite{Saeedia,Zhai}.

Quantum interference between the two arms of the ring provides suitable
means for controlling the spin in the nano-scale, which has been
proven by the Green's function method \cite{Wang,Sun1} or Griffith's
boundary conditions \cite{Griffith,AC1,Bellucci,Citro,Tang,Saeedia,Zhai}. The
first order linear approximation with full transparent contacts was also
reported \cite{Nitta,AC2,Gumber,Li}, albeit without the backscattering effect.
The S-matrix method \cite{Hatano,Naeimi} presents a
rough assessment of the backscattering by fixing the energy-dependent
coupling parameter between the leads and ring as constant.
We note that previous works on spin transport properties in the
quantum ring were not comprehensive. For spin-unpolarized input current
these works often only focused on spin polarization in the $z$ direction or
the direction of the eigenstates of the ring. The total polarizability,
polarization direction, and spin polarization in arbitrary directions
were rarely discussed. Work in the case of the arbitrarily spin-polarized
incident are difficult to find in the literature.

In this work, we present an analytical model for one-dimensional (1D)
rings and numerical studies of realistic two-dimensional (2D) quantum rings
in the non-equilibrium Green's function (NEGF) method \cite{chang} where
both the Rashba and Dresselhaus SOCs are present. We derive the formula
for the transmission rates for arbitrary spin polarization and
generalize them to the cases of the polarized incident spin. A density matrix
describing the spin-polarized (in arbitrary direction) input current is also
introduced into the Green's function equation, which results in the same
results obtained by the analytical 1D model.
%This density matrix can also be
%a mixed state describing a partial polarized input current.
%We study in detail how the spin and charge transport occurs through the quantum
%ring device. If the input current is spin-unpolarized, the total
%polarizability, polarization direction, and spin polarization in arbitrary
%directions are discussed.
The transmission rate $T$ can be up to unity with the fully polarized output in
a proper magnetic field and with a proper Rashba SOC.

When the input current is spin-polarized the transmission rate depends on the
direction of the input polarization and the output current is still spin
polarized. So the quantum ring is also acting as a spin torque which may be
useful in spintronics. This property also guides us finding the way to design 
an omnidirectional spin analyzer. In contrast, the optical polarization 
analyzer is simpler since the polarization is perpendicular to the direction 
of the light. However, the spin polarization can be along an arbitrary 
direction on the Bloch sphere. The spin analyzer in a particular direction can 
be achieved in the ferromagnetism systems \cite{handbook}. The arbitrary spin 
analyzer needs the light involved \cite{huang,jozwiak}, which is difficult to 
be integrated. Here, we just need to measure the conductances in different 
strengthes of the SOC to obtain the polarization of the incident spin, which 
is easier to integrate on the chip. It is interesting that in such a simple 
system, the spin filter, spin polarizer and spin analyzer can be achieved by 
just tuning the magnetic field or the Rashba SOC via the 
gate \cite{Rash01,Rash03,Rash04,Rash05}.

%The manuscript is organized as following. In Section II, we write down the
%tight-binding Hamiltonian of the quantum ring with Rashba and Dresselhaus SOCs.
%Then introduce the Green's function method in calculating the transport
%properties. In Section III, we calculate the transport properties in the
%one-dimensional model analytically. The AB and the AC effects are considered
%fully, so that the spin filtering, polarizing effects can be obtained
%quantitatively comparing with the numerical results. Moreover, we calculate
%the transmission rates of the polarized incident spin. In Section IV, we
%explore how the unpolarized spin current is polarized by the quantum ring
%device numerically in the nonequilibrium Green's function method. We then
%propose an all-electric spin analyzer in Sec. V by measuring the transmission
%rates of the polarized incident spin in three different SOCs. Finally, we
%conclude our work in the last section.

\section{The transport properties in the one-dimensional model}\label{1dmodel}

To understand the transport properties in a quantum ring, the one-dimensional
(1D) model is usually applied. The ring is contacted with the left and the
right leads at $\varphi=\pi$ and $0$, respectively. In this work, we suppose
that the electron is injected from the left lead, then it travels
through the ring in two different paths, one from $\varphi=\pi$ to $0$
clockwise (the upper arm) and the other from $\varphi=\pi$ to $2\pi$
counterclockwise (the lower arm), as shown in Fig. \ref{fig1}(a).

As discussed in the previous work \cite{peng}, the 1D model works very well
when the radius is not too large. The 1D model here, at least, is a good
approximation which results in the correct physical pictures. Another
approximation of neglecting the Zeeman effect is also adopted. In the
relatively low magnetic field ($B <3 $T), the Zeeman coupling is weak
and could be neglected. We can also numerically verify that this
approximation is appropriate in low magnetic fields.

%\subsection{The Rashba SOC in the ring}\label{1drashba}

For simplicity, we first consider only the Rashba SOC being present.
If the Zeeman coupling is neglected the energy spectrum of the 1D ring
is given by \cite{sheng,AC1,AC2,AC3,AC4}
$E_n^{\mu}=\tau\left( n_j^{\mu}-\frac{\Phi_{AB}}{2\pi}-\frac{\Phi
_{AC}^{\mu}}{2\pi}\right)^2$
where $n_j^{\mu}$ is the orbital quantum number, and the index $\mu=1,2$
represents the spin eigenstates $\left\vert \uparrow\right\rangle$ and
$\left\vert\downarrow\right\rangle$, and $j=\pm$ represents the clockwise
and counterclockwise electron motions, respectively. Also, $\tau=\frac{\hbar^2}
{2m^* r_0^2}$ is the energy unit, $\Phi_{AB}=2\pi N$ is the AB phase with the
relative magnetic flux $N=\frac{eBr_0^2}{2\hbar}$, and $\Phi_{AC}^{\mu}=
(-1)^{\mu}\left(\sqrt{1+4\beta_{1}^2}-1\right)\pi$ is
the AC phase with $\beta_1= g_1 m^* r_0 /\hbar$.

The corresponding eigenstates are given by $\Psi_j^{\mu}\left(\varphi
\right)=\frac1{\sqrt{2\pi}}e^{-in_j^\mu\varphi}\chi^{\mu}\left(\varphi
\right)$, where $\chi^1(\varphi)=\left(\cos\frac{\theta_1}2,-e^{i\varphi}
\sin\frac{\theta_1}2 \right)^T$ and $\chi^2 (\varphi)=\left(\sin\frac
{\theta_1}2,e^{i\varphi}\cos\frac{\theta_1}2 \right)^T$, with $\tan
\theta_1=2 \beta_1$ \cite{AC4}. It is clear that the directions of the spin
polarization are along $(\theta_1,\varphi)$ and $(\pi-\theta_1,\pi+\varphi)$
for the two eigenstates respectively.

The schematic diagram of the total transport is explicitly drawn in Fig.
\ref{fig1}(a). The incident current can be decomposed into the two eigenstates
$\chi^{1,2}$, and the electron is transported by these two channels. The
transmission rate is given by (see the Method),
\begin{equation}
T_{\mu}=\frac{K \Phi^\mu}{K K'+\left[4k_0^2\left(\Phi^\mu-K' \right)
+k^2\sin^2(\pi k_0r_0) \right]^2 },
\label{T1}
\end{equation}
where $K=16k^2k_0^2\sin^2(\pi k_0r_0)$, $\Phi^\mu=\cos^2\frac
{\Phi_{AB}+\Phi_{AC}^{\mu}}2$ and $K'=\cos^2(\pi k_0r_0)$.
For vanishing magnetic field the AB phase vanishes and the SOC induced
energy shift $U_0$ is neglected, then Eq.~(\ref{T1}) agrees with the results
obtained in Ref.~\cite{AC1,AC3}. If there is a constant potential $U$ added at
the contact then the magnetic field for $T_\mu=0$ is not changed while the
magnetic field for $T_\mu=1$ is slightly shifted. Hence, for the spin filter
the contact defect is not very important.

The numerator of Eq.~(\ref{T1}) indicates that
the transmission rate oscillates with the incident energy $E$, and $\cos^2
\frac{\Phi_{AB}+\Phi _{AC}^{\mu}}2$ means that $T_{\mu}$ oscillates with
the increase of the magnetic field $B$ or the coupling strength of the SOC
$g_1$. When the magnetic field vanishes, $\Phi_{AB}=0$ and $T_1=T_2$,
resulting in a completely unpolarized transport if the incident spin is
unpolarized. However, if both of the AB and the AC phases are taken into
consideration in a proper magnetic field the spin can be fully polarized
after traversing the ring.

If we want an $100\%$ polarized spin current output then the phases must
satisfy $\Phi_{AB}+\Phi_{AC}^\mu= \pi$ so that the eigenstate $\chi^\mu$ is
completed filtered out, and only the other eigenstate is left. For a given SOC
different AB phases (different magnetic flux) lead to different eigenstate
filtering. The magnetic flux difference of the two nearest eigenstates
filtering is then given by $\Delta N= \frac12 (\sqrt{1+4 \beta_1^2}-1)$.
This result of constructing a perfect spin filter is consistent
with the results of the S-matrix method \cite{Hatano}.

If the incident spin is unpolarized, then the spin can be composed of an
arbitrary direction $(\theta',\varphi')$ and its opposite direction $(\pi-
\theta', \pi+\varphi')$ independently. The two transport channels do not
interfere with each other, and the transmission rates can be obtained by
projecting the two eigen transmission rates onto the two directions,
$
T_{\left(  \theta',\varphi'\right)  +}  =\sum_{\mu} \left| \left[ \chi^
{\left(  \theta',\varphi'\right)  } \right]^\dag \chi^{\mu}\left(  0\right)
\right| ^{2}T_{\mu}$, and $
T_{\left(  \theta',\varphi'\right)  -}  =\sum_{\mu} \left| \left[\chi^
{\left( \pi-\theta',\pi+\varphi'\right)  } \right]^\dag \chi^{\mu}\left(
0\right) \right| ^{2}T_{\mu},
$
where $\chi^{(\theta',\varphi' )}\equiv\left( \cos\frac{\theta'}{2},
e^{i\varphi'}\sin\frac{\theta'}{2}\right) ^T$. The upper index of $\chi$
stands for the direction of the spin of the state. The spin polarization of
the outcoming current in an arbitrary direction can be found to be
\begin{equation}
P_{\left(  \theta',\varphi'\right) }= \left[\chi^{1}\left(
0\right)  \right]^\dag \sigma_{\left(  \theta',\varphi'\right) }
\chi^{1}\left(  0\right) P_{\chi} \label{P0},
\end{equation}
where the spin matrix along the direction of $(\theta', \varphi')$ is
$
\sigma_
{\left(  \theta',\varphi'\right)  }=(\sigma_{x}\cos\varphi'+\sigma_{y}\sin
\varphi')\sin\theta'+\sigma_{z}\cos\theta',
$
and $P_\chi=(T_1-T_2)/(T_1+T_2)$
is the spin polarization in the direction of the two eigenstates at the
contact, $(\theta_1/2, 0)$. Since $|P_{\left(  \theta',\varphi'\right)} |\leq
|P_\chi|$, the outcoming polarization is always along the direction of the
eigenstate $\chi^{1}$ or $\chi^2$.

The transmission rates when the incident spin is unpolarized are well studied.
Next we consider the case where the incident spin is polarized in an arbitrary
direction along $(\theta, \varphi)$. Irrespective of the incident electron
is a pure or a mixed state, the transmission rate is always obtained by
\begin{eqnarray}
T^{\left(  \theta,\varphi\right) }  &=&\sum_{\mu}\left\vert \left(\chi^{\left(
\theta,\varphi\right)  } \right)^\dag \chi^{\mu}\left(  \pi \right)
\right\vert ^{2}T_{\mu},  \notag \\
&=& T_{1}\cos^{2}\left(\frac{\theta_{\Delta}^{in}}{2} \right) +T_{2}\sin^{2}
\left( \frac{\theta_{\Delta }^{in}}{2} \right), \label{TT1}
\end{eqnarray}
where $\theta^{in}_{\Delta}$ is the angle between the direction $(\theta,
\varphi)$ and the direction of the spin polarization of $\chi^1 (\pi)$ which
is $(\theta_1, 0)$. It means that the arbitrarily polarized spin is
projected to the two conjugate eigensates of the ring, and the transmission
rate of the spin is the sum of the two eigen channels. Moreover, for the
unpolarized incident current, we can decompose it into two conjugate parts,
and we get $T^{\left( \theta,\varphi\right)} +T^{\left( \pi-
\theta,\pi+\varphi\right) } =T_1+T_2$.

In fact, we can define the transmission rate $T^{\left( \theta,\varphi\right)}
_{\left(  \theta',\varphi'\right)\pm}$ where the upper index is the
polarization of the incident spin and the lower index represents the
transmission rate along the direction $\left(  \theta',\varphi'\right)$ (for
$\left(  \theta',\varphi'\right)+$) or $\left( \pi-\theta',\pi+\varphi'\right)$
(for $\left(  \theta',\varphi'\right)-$). If the incident electrons
are being injected one by one and is supposed to be a pure state, then the outcoming
wave function at the right lead can be found as
$
\chi_{out}^{(\theta,\varphi)}=\sum_{\mu} \left[\chi^{\mu}\left(  \pi\right)
\right]^\dag  \chi^{\left(  \theta,\varphi\right)}  t_{\mu} \chi^{\mu}\left(
0\right).
$
The transmission and the polarization rates are then given by
\begin{eqnarray}
T_{\left(  \theta',\varphi'\right)  \pm}^{\left(  \theta,\varphi\right) } &=&
\left[\chi_{out}^{(\theta,\varphi)}\right]^\dag \sigma_{\left( \theta,\varphi
\right) \pm} \chi_{out}^{(\theta,\varphi)} \label{TP} \\
P_{\left(  \theta',\varphi'\right) }^{\left(  \theta,\varphi\right) } &=&
\left[\chi_{out}^{(\theta,\varphi)}\right]^\dag \sigma_{\left( \theta,\varphi
\right) } \chi_{out}^{(\theta,\varphi)}, \label{torque}
\end{eqnarray}
where $\sigma_{\left( \theta,\varphi\right) \pm} = |(\theta,\varphi)\pm\rangle
\langle (\theta,\varphi)\pm|$ is the density matrix of the eigenstate of the
matrix $\sigma_{(\theta,\varphi)}$.
%It is obviously that $T_{\left(  \theta',\varphi'\right) }^{(\theta,\varphi)}
%=T_{\left(  \theta',\varphi'\right) +}^{\left(  \theta,\varphi
%\right)}-T_{\left( \theta',\varphi'\right) -}^{\left(  \theta,\varphi\right)}$
%and $T^{\left(  \theta,
%\varphi\right)}=T_{\left(  \theta',\varphi'\right) +}^{\left(  \theta,\varphi
%\right)}+T_{\left( \theta',\varphi'\right) -}^{\left(  \theta,\varphi\right)}$.
%The spin polarizability along this direction is
%$ P^{(\theta,\varphi)}_{\left( \theta',\varphi'\right) }=T_{\left(  \theta',
%\varphi'\right) }^{(\theta,\varphi)}/ T^{\left(  \theta, \varphi\right)}$.

By analyzing Eq. (\ref{TT1}), it is easy to obtain that $\max(T^{\left(
\theta,\varphi\right) })=\max(T_1, T_2)$ and $\min(T^{\left( \theta,\varphi
\right) })=\min(T_1, T_2)$. Therefore, the incident spin having maximum and
the minimum transmission rates must be parallel to the polarization directions 
of the two eigenstates, respectively.

The generic spin torquing is given by Eq. (\ref{torque}), but the presence of
a magnetic field makes the analytical result a bit complicated. For
simplicity, we consider the magnetic field approaching zero, so that
$\Phi_{AB}\rightarrow 0$ and $T_1 = T_2$. The spin polarizations for an
arbitrarily polarized incident current are
\begin{equation}
\left\{
\begin{array}{c}
P_{x}^{\left(  \theta ,\varphi \right) }  =-\sin 2\theta_1 \cos\theta
+\cos2\theta_1 \sin\theta\cos\varphi, \\
P_{y}^{\left(  \theta ,\varphi \right) }  =\sin\varphi\sin\theta, \\
P_{z}^{\left(  \theta ,\varphi \right) }  =\cos2\theta_1 \cos\theta
+\sin2\theta_1 \sin\theta \cos\varphi .
\end{array}
\right. \label{pxyz}
\end{equation}
When $\varphi=0$, then $P_{x }^{\left(  \theta ,\varphi \right)  }  =-\sin
\left(2\theta_1-\theta \right), P_{y }^{\left(  \theta ,\varphi \right)  } =0,
P_{z}^{\left(  \theta ,\varphi \right)  } =\cos\left( 2\theta_1-\theta \right)$.
It means that the incident and outcoming spins are all in the $xOz$ plane,
the spin passes the ring and is torqued a fixed angle in the $xOz$ plane,
$\left(\theta_{out},\varphi_{out}\right) =\left( \theta-2\theta_1, 0\right)$.
It can be intuitively understood by the spin textures of the eigenstates
that there is no $y$ component spin at $\varphi=0,\pi$ in the ring
\cite{peng}. The torqued angle is only related to the strength of the SOC.
This special case goes back to the result obtained in Ref.~\cite{Shen}, and
the more special case, $P_{z}^{\left(0,0\right)}=\cos\left(2\theta_{1}\right)$
was obtained in the path-integral approach \citep{Lucignano}. A series
of the ring may be able to tune the spin polarization arbitrarily.

%\subsection{The Dresselhaus SOC and the case of two SOCs combined}

If only the Dresselhaus SOC is present, the analysis above
is still valid, but some terms need to be changed. The AC phase needs to be
replaced by $\Phi^{\mu}_{AC}=-(-1)^\mu\left( \sqrt{1+4 \beta^2_2}\right)\pi$
where $\beta_2=g_2 m^* r_0 /\hbar$. The eigenstates of the ring also need
to be changed to $\chi^{1}(\varphi)=\left(  \cos\frac{\theta_2}{2},ie^{-i
\varphi} \sin \frac{\theta_2}{2}\right)^{T},\ \chi^{2}(\varphi)=\left( \sin
\frac{\theta_2}{2}, -e^{-i\varphi}\cos \frac{\theta_2}{2}\right) ^{T}$, with
$\tan \theta_2= 2\beta_2$, and the additional potential is $U_0=-\beta^2_2$.
All other calculations remain unchanged.

When both the SOCs are present
then it would be difficult to have analytical results for the transport
problem. We then seek the solutions numerically in the NEGF method.

\section{Numerical results of the Spin and charge transport properties in
two-dimensional models}

\begin{figure*}[ptb]%
\centering
\includegraphics[
scale=0.8]%
{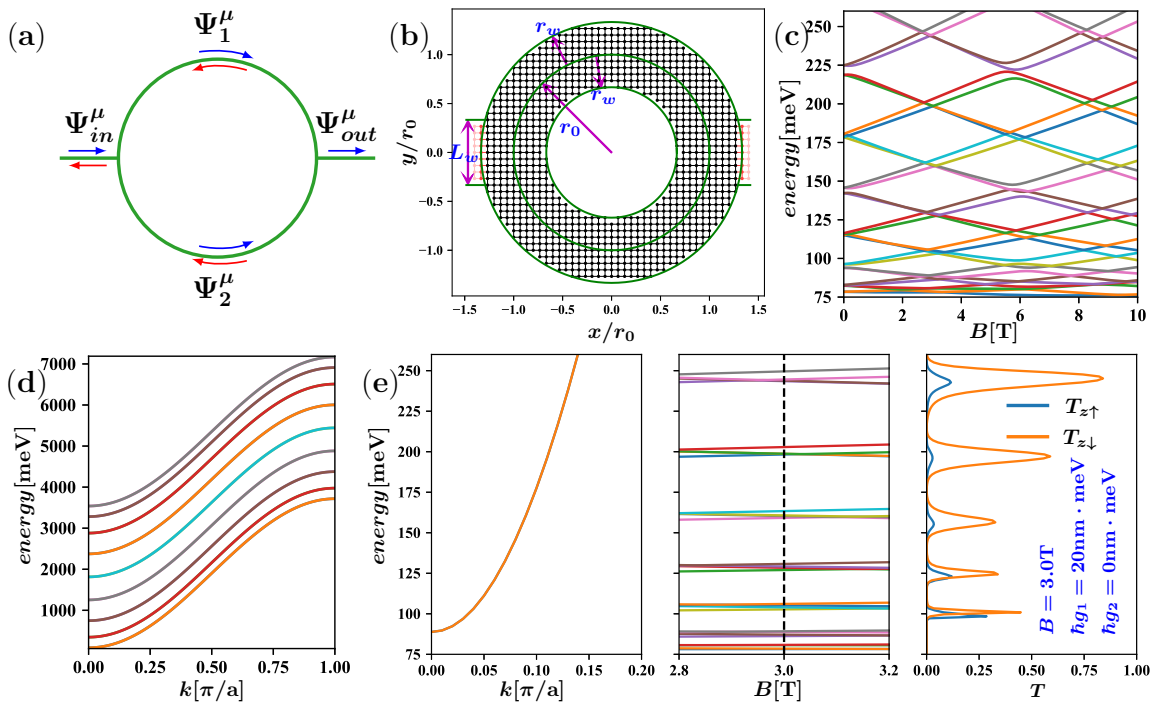}%
\caption{(color online). (a) The schematic transport in a 1D ring. (b) The
device with $r_{0}$=15 nm, $r_{w}$=5 nm, lead (red) width $L_{w}$=10 nm and
lattice constant $a$=1 nm. (c) The energy spectrum of tight-biding ring
(black) with $\hbar g_1$ = 20.0 nm$\cdot$ meV and $g_2$ = 0. (d) The energy
spectrum of the lead without the Zeeman coupling or SOCs. (e) Three figures
from left to right: The energy of the input electrons; Around $B=3$T, the
energy spectrum of the ring within the energy of the input electrons, i.e.
$E<250$ meV; The transmission versus the energy of the input electrons.}
\label{fig1}
\end{figure*}
%EndExpansion

The spin transmission rates $T_\alpha$ and the spin polarization rates
$P_\alpha$ are important variables characterizing the transport properties.
%see more details of the formalism in the appendix.
The spin polarization rate $P_\alpha$ is the probability of the
spin of the outcoming electron projected to the $\alpha$ axis. $P_0$ is the total
polarization of the outcoming spin. If $P_0=1$, then the spin of the current
at the drain is fully polarized along a certain direction, otherwise the
outcoming current contains different components of the spin at the same time
and it is not fully polarized.
We here numerically calculate the two rates to explore the transport properties
of a more realistic two-dimensional quantum ring contacted by the source and
drain on the two ends of a diameter. We then discuss how the spin of the
current is polarized and filtered by the quantum ring with the SOCs when the
incident electron is spin unpolarized, and compare the 2D numerical results
with the analysis in the 1D model.

For simplicity and without loss of generality we consider the ring on the 
surface of the InAs semiconductor. We adopt the tight-binding Hamiltonian 
(details shown in the appendix) to perform the
numerical calculations by applying the Green's functions \cite{dattabook,chang}.
The device is indicated in Fig. \ref{fig1}(b), in which the lattice constant
is $1$ nm \cite{latticeconstant}. The energy spectrum of the ring without the
source and the drain in such a tight-binding model is similar to that of the
ring in the parabolic potential calculated in Fock-Darwin basis \cite{QD_book},
as shown in Fig. \ref{fig1}(c). So the tight-binding model itself is reliable
and is a very accurate approximation for the real physical system.

Fig.~\ref{fig1} shows the two-lead transport device and an example of the
transport property of the ring. The lead is $10$ nm wide in the $y$ direction
and is semi-infinite along the $x$ axis. We consider low-energy transport only,
and the input electrons are on the lowest energy band allowed in the lead as
shown in Fig.~\ref{fig1}(d). In Fig.~\ref{fig1}(e), we find that the
transmissions are only allowed when the energy of the incident electron is
close to the energy levels of the ring.

Previous studies have offered the possibilities that the quantum ring with SOCs
can act as the spin filter and polarizer. We apply the NEGF method in a full
2D model quantum ring with SOCs, as shown in Fig. \ref{fig1}. Moreover, we
take all the realistic conditions, Zeeman coupling, finite width of the ring
and the rotational symmetry breaking, into consideration. In fact, the 1D model
still works qualitatively. The spin filtering can be basically
explained by the transmission rates $T_\mu$ in Eq. (\ref{T1}) varying with the
competition of the AB and the AC phases in different magnetic fields. In a
proper magnetic field and with a proper SOC, one channel can be shielded and
the other one is fully survived, so that both $T_0$ and $P_0$ can be up to 1.
Details can be found in the appendix.

As discussed in the 1D model, if only the Rashba SOC is present
then $T_y$ and $P_y$ will be suppressed. The direction of the spin polarizer
can be tuned by the strength of the Rashba SOC in the plane $xOz$. If only the
Dresselhaus SOC is present, then $T_x$ and $P_x$ will be suppressed. The
direction of the spin
polarizer is then in the plane $yOz$. If both of the SOCs are present then
the situation becomes complex and the spin polarizer can be controled more
widely. However, we find that if the outcoming spin needs to be polarized
well, then it is better to keep one SOC dominating the system. The
competition of the two SOCs makes the spin more difficult to be polarized,
as shown in the appendix.

We note that in such a simple device the spin filtering and spin polarizer
can be realized. The unpolarized spin is transported through the simple
quantum ring with Rashba SOC, and then the outcoming spin is polarized. The
directions other than the outcoming polarization are filtered, and the current
is spin polarized. Moreover, the Rashba SOC can be easily tuned, so that the
polarization of the outcoming spin can be easily tuned by a gate.

\section{Isotropic all-electric Spin analyzer}

%We have studied how the quantum ring polarizes the
%spin of the incident electron with the help of the SOCs.
Now we would like to
consider the case when the incident electrons are already fully polarized.
Similar to the light polarizer, the ring with the SOCs in fact can be acted
as a spin analyzer. If the incident electron is already spin polarized
in the direction of $(\theta_{in},\varphi_{in})$ in the spherical coordinate
of the spin space, then the transmission rate is given by
\begin{equation}
%T_{\alpha}(E)=\text{Tr} \{\sigma_{\alpha}[ \Gamma_{R}(E) G_{RL}(E)
%\sigma_(\theta_{in},\varphi_{in}) \Gamma_{L}(E) G_{LR}^{\dag}(E)]\},
T_{\alpha}(E)=\text{Tr} \left[ \sigma_{\alpha}\Gamma(E) G(E)
\sigma_{(\theta_{in},\varphi_{in})+} \Gamma(E) G^{\dag}(E) \right],
\label{polarizedtrans}
\end{equation}
where $\sigma_{(\theta_{in},\varphi_{in})+}$ is the density matrix of the
polarized state,% defined in Sec. \ref{1dmodel},
the Green's function $G$ and
the broadening function $\Gamma$ can be found in Ref.~\cite{dattabook}. We
note that the outcoming spin is still spin polarized, but is torqued by an
angle given by Eq. (\ref{pxyz}).
%The general spin torque angle can be found in App. \ref{app2}.

Using Eq.~(\ref{polarizedtrans}), we can clearly
decompose the unpolarized incident $\psi^{in}$ in the basis of $\sigma_z$. In
the density matrix form, $|\psi^{in}\rangle \langle \psi^{in}|= \left(
|\psi^{in}_{z\uparrow}\rangle \langle \psi^{in}_{z\uparrow}|+ |\psi^{in}
_{z\downarrow}\rangle \langle \psi^{in}_{z\downarrow}| \right)/ 2$.
The incident wave function can be divided into two parts with opposite spin
polarization, and each part provides a transport channel. The total
transmission rate is the sum of the transmission rates of the two channels,
since there is no coherence between the two channels.
%In Appendix \ref{app3},
In the appendix, we can clearly see how the spin textures and 
the current evolve in the ring
for different channels in which the spin is decomposed along $+z$ or $-z$.

%The unpolarized incident electron with $E_{in}=198.5$ meV has
%the spin transmission rates $T_0=0.186$ and $(T_x, T_y, T_z)= (-0.078,0.002,
%0.169)$ and the polarization $P_0 \approx 1$, so that the outcoming electron
%is spin polarized in the direction of $(\theta_{out},\varphi_{out})=(0.138\pi,
%0.993\pi)$, which is essentially the eigenstate of the ring $\chi^1 (0)$
%discussed in Section \ref{1dmodel}. The charge transmission is obtained by
%the average of the two channels shown in Figs. \ref{figapp4}(a) and (b),
%since the unpolarized states can be decomposed to the two channels equally.

\subsection{Transmission rates for the polarized incident spin current}

%We note that when the incident spin is polarized in a special direction
%$(\theta_{in}^{max},\varphi_{in}^{max})=(0.138 \pi, 0.007 \pi)$ which is
%close to $\chi^1(\pi)$ and is the mirror symmetric of the outcoming
%spin $(\theta_{out}, \varphi_{out}) = (0.138\pi, 0.993\pi)$ to the plane
%$yOz$, the transmission rate is maximum.
%The outcoming polarization angle is somehow related to the input angle in
%Figs. \ref{fig3}(b) and (c). However, we note that when the transmission
%rate is sufficiently large (the red region in Fig. \ref{fig3}(a)), the
%outcoming angle is almost fixed to $(\theta_{out}, \varphi_{out}) =
%(0.138\pi, 0.993\pi)$. Otherwise, the outcoming angle is changed when the
%transmission rate $T$ is very small.

\begin{figure*}[ptbh]%
\centering
\includegraphics[
height=2.60in,
width=6.70in
]%
{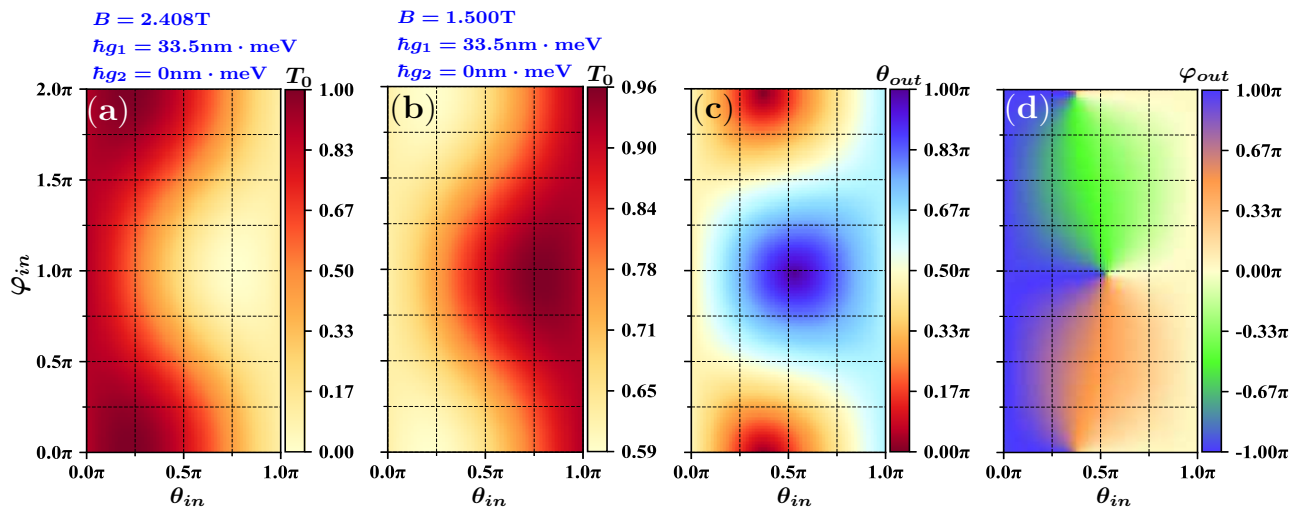}%
\caption{(color online). The energy of the incident electron is $E_{in}
=198.5$ meV. (a) The total transmission rate $T_{0}$ for different angles of
the polarization of the incident electron $(\theta_{in},\varphi_{in})$ when
$\hbar g_{1}=33.5$ nm$\cdot$ meV and $g_{2}=0$ at $B=2.408$ T. (b) The total
transmission rate $T_{0}$ and (c)-(d) the angles of the polarization of the
outgoing electron $(\theta_{out},\varphi_{out})$ for different polarization
of the incident electron $(\theta_{in},\varphi_{in})$ at $B=1.5$ T.}
\label{fig2}
\end{figure*}

We suppose that the ring is only coupled by the Rashba spin-orbit interaction
$\hbar g_1=20$ nm$\cdot$ meV and the incident current is already spin polarized.
The polarization direction of the incident spin $(\theta_{in},\varphi_{in})$
varies and the charge transmission rate is indicated in Fig. \ref{fig2}.
%The maximum rates are $T_0^{max}=0.372$ and $(T_x, T_y, T_z)=
%(-0.156,0.004,0.337)$, which are twice of the rates of the unpolarized
%incident spin.
Fig. \ref{fig2}(a) shows the case when $T_{1}=1, T_{2}=0$ and $P_0=1$.
The one-dimensional analytical model predicts that the outcoming polarization
is along the eigenstate $\chi^{1}(0)$, $(0.161\pi,\pi)$, and the $\chi^2$
channel is closed $T_2=0$. In the two-dimensional model, it indicates that
the outcoming polarization is along $(\theta_{out},\varphi_{out}) = (0.214
\pi,0.984\pi)$ always, where the channel of $\chi^1$ is free to transport and
the other channel ($\chi^2$) is completely closed. It means that the
polarization angle of the eigenstate of the 2D ring at $\varphi=0$ is
$(0.214\pi,0.984\pi)$. This difference comes from the Zeeman effect and the
finite width. It implies that these effects can also generate a finite $P_y$
in the ring with the Rashba SOC only, which is {\it significantly differenti}
from the 1D model.

Moreover, the incident polarization with the maximum transmission rate among
all the directions in the spin space is also along the eigenstate $\chi^1_{2D}
(\pi)$, $(\theta_{in}^{max},\varphi_{in}^{max})=(0.214\pi,0.016\pi)$. We note
that $(\theta_{in}^{max},\varphi_{in}^{max})$ and $(\theta_{out},
\varphi_{out})$ are mirror symmetry to the $z$ axis.

The transmission of the spin-polarized input current in arbitrary direction is
determined by the projection of $(\theta_{in},\varphi_{in})$ to $(\theta
_{in}^{max},\varphi_{in}^{max})$, since the channel of $\chi^2_{2D}$ is closed.
For a more general case, both transport channels of the eigensates allow
electrons to pass ($T_{0}^{\max},T_{0}^{\min} >0$), as shown in Figs.
\ref{fig2}(b) and (c), the maximum transmission rate $T_{0}^{\max}=0.962$
corresponds to the incident polarization $(\theta_{in}^{max},\varphi_{in}
^{max})=(0.792\pi,0.963\pi)$, and its output polarization is $(\theta_{out}
^{max},\varphi_{out}^{max})=(0.792\pi,0.037\pi)$. For the minimum transmission
rate, we have $T_{0}^{\min}=0.591$, $(\theta_{in}^{\min},\varphi_{in}^{\min})
=(0.208\pi,1.963\pi)$, $(\theta_{out}^{\min},\varphi_{out}^{\min})=(0.792\pi,
1.037\pi)$. In this case, $(\theta_{out},\varphi_{out})$ is no longer a fixed
angle along $\chi^1_{2D}$, but changes with the angle of
incidence $(\theta_{in},\varphi_{in})$.

%It is worth noting that the following relationships have not changed in the
%numerical results:
%$(\theta_{in\left(  out\right)  }^{max},\varphi_{in\left(  out\right)  }%
%^{max})=(\pi-\theta_{in\left(  out\right)  }^{\min},\pi+\varphi_{in\left(
%out\right)  }^{\min})$, $(\theta_{in}^{max\left(  \min\right)  },\varphi
%_{in}^{max\left(  \min\right)  })=(\theta_{out}^{max\left(  \min\right)  }%
%,\pi-\varphi_{out}^{max\left(  \min\right)  })$. It is driven by the spin
%textures of the system and is topologically protected \cite{peng}.

Interestingly, we also find numerically that the general relation between
the incident angle and the charge transmission rates $T_0$ is given by
\begin{equation}
T_{0} =T_{0}^{\max}\cos^{2}\left(  \frac {\theta_{\Delta}}{2}\right)
+T_{0}^{\min}\sin^{2}\left(\frac{\theta_{\Delta}}{2}\right), \label{tpj}
\end{equation}
where $\theta_{\Delta}$ is the angle between the incident spin
polarization $(\theta_{in},\varphi_{in})$ and the special angle $(\theta_{in}
^{max},\varphi_{in}^{max})$, whether the outcoming spin is polarized or not.
This equation is exactly the same as Eq.~(\ref{TT1}) that we found for
the 1D model. The only difference is that in Eq.
(\ref{TT1}), $T_{1,2}$ correspond to the transmission rate of the eigenstates
of the ring $\chi^{1,2}$. However, in the 2D ring the $T_0^{max}$ and
$T_0^{min}$ correspond to the eigenstates of the 2D ring which are a little
different from those of the 1D ring. The arbitrary spin is projected to the
angles of the eigenstates of the ring, $(\theta_{in}^{max},\varphi_{in}^{max})
$ and $(\pi- \theta_{in}^{max},\varphi_{in}^{max}+\pi)$. This projection
then gives directly the transmission rate in Eqs. (\ref{TT1}) and (\ref{tpj}).
The Zeeman coupling, circle symmetry breaking, and finite width only change
the spin-polarization direction of eigenstates $\chi^{\mu}$, the properties
predicted by the 1D analytical model are retained, which implies that we
could use the quantum ring to design the integrable spin devices.

\subsection{Design of a spin analyzer}

The ring acts as a spin torque: it allows the electron to pass but the
spin polarization must be torqued. If the ring is coupled by the Dresselhaus
spin-orbit interaction only, the similar spin torque occurs. The only
difference is the outcoming angle of the spin, which is the mirror symmetry
of the incident angle (for the maximum transmission rate only) to the plane
$xOz$. The direction dependent transmission rate is also given by Eqs.
(\ref{TT1}) and (\ref{tpj}).

According to the property of the angle dependent transmission rate in Eq.
(\ref{tpj}), we can realize a spin analyzer in the ring device. Before the
measurement, we need to know $T_0^{max}$ and $T_0^{min}$ in a given magnetic
field. They can be determined by the measurement of the transmission rates of
the known spin polarized incidents. We use three spin polarized incident with
$P_{x,y,z}=1$, respectively, and one spin unpolarized incident to identify
the following parameters: $T_0^{max},T_0^{min},\theta_{in}^{max},\varphi_{in}
^{max}$. The transmission rate for the unpolarized incident is marked as $T$, and
we have already known $T=T_0^{max}+T_0^{min}$. The three transmission rates
for different spin polarization incident are marked as $T(x), T(y)$ and $T(z)$.
Applying Eq. (\ref{tpj}) to $T(x), T(y)$ and $T(z)$, we find another three
equations. So four equations in all can be solved and the four variables
$T_0^{max},T_0^{min},\theta_{in}^{max},\varphi_{in}^{max}$ are found.

\begin{figure*}[ptb]%
\centering
\includegraphics[
scale=0.7
]
{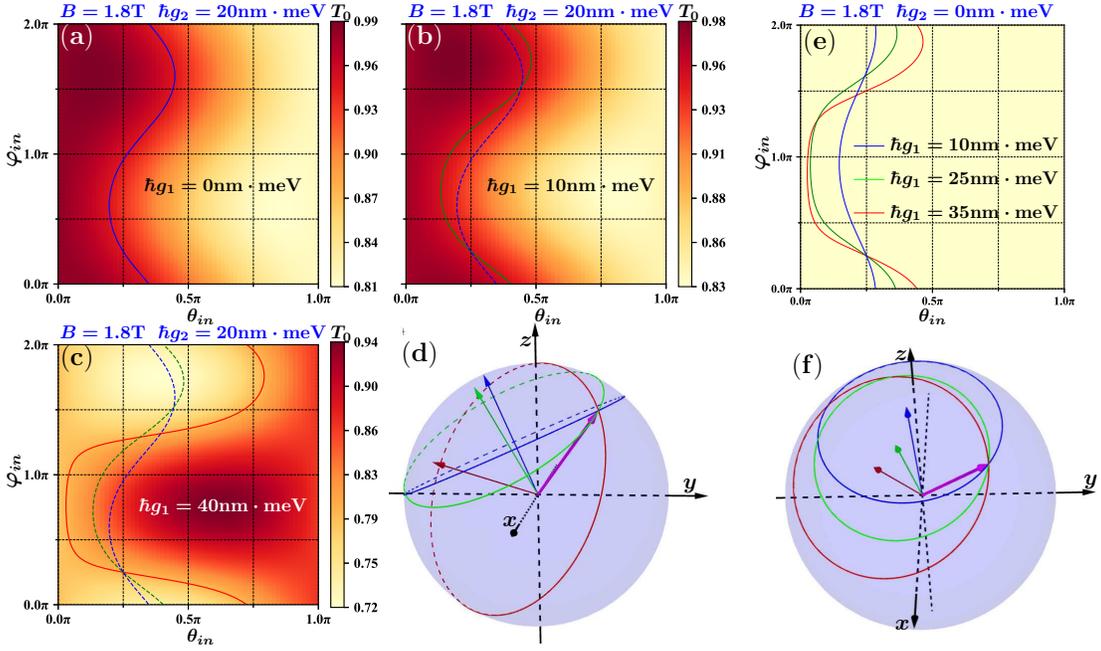}
\caption{(color online). The energy of the incident electron is $E_{in}%
=198.5$ meV, and the background magnetic field $B=1.8$ T. (a)-(d) The colors
represent the transmission rates for the SOCs $\hbar(g_{1},g_{2}%
)=(0,20),(10,20),(40,20)$ nm$\cdot$ meV, respectively. The solid curve in (a)
- (d) represents the possible angles of the incident polarization after the
first, second and the third measurements, respectively. The dash lines in (b)
and (c) represents the previous measurements. (d) The possible polarization
is a cone for each measurement in the spin space. The blue, green and red
arrows stand for the vector $(\theta^{max}_{in},\varphi^{max}_{in})_{1,2,3}$,
respectively. The intersection of the three cones is the purple vector
representing the incident polarization. (e)-(f) The intersections of the
possible angles in three Rashba SOCs $\hbar g_{1}=10,25,35$ nm$\cdot$ meV,
respectively, when the Dresselhaus is absent.}
\label{fig3}
\end{figure*}
The scenario to analyze the spin polarization by detecting
the charge transmission rates for different SOCs then can be established.
The scheme is described as follows:

First, the ring is coupled by the two spin-orbit interactions $(g_1,g_2)$.
The transmission rates are shown in colors in Fig. \ref{fig3}(a).
$T_0^{max}$ and the corresponding incident polarization $(\theta_{in}^{max},
\varphi_{in}^{max})_1$ is already known, as the blue vector in Fig.
\ref{fig3}(d). Once we measure the transmission rate $T_0$, we can find the
angle $\theta_{\Delta1}$ between the incident
polarization angle $(\theta_{in} ,\varphi_{in})$ and $(\theta_{in}^{max},
\varphi_{in}^{max})_1$ by applying Eq. (\ref{tpj}). However, the possible
polarization direction in the three-dimensional space of the spin can be along
any element of the cone, shown as the blue circle in Fig. \ref{fig3}(d).
we project the angles of the elements of the cone onto the $(\theta, \varphi)$
plane to obtain the solid line in Fig. \ref{fig3}(a).

Second, we tune the Rashba SOC and the transmission rates are shown in
colors in Figs. \ref{fig3}(b). The angle of the maximum transmission rate,
$(\theta_{in}^{max},\varphi_{in}^{max})_2$, is represented by the green vector
in Fig. \ref{fig3}(d). Then measure the transmission rate to obtain the angle
$\theta_{\Delta2}$ to find the second cone. The spin polarization is possibly
located in the solid green line in Fig. \ref{fig3}(b), where the dashed line
represents the first measurement. So the incident polarization must be at one
of the intersection points of the two lines.

Thirdly, we tune the Rashba SOC again to find the third line which is shown in
Fig. \ref{fig3}(c). The three lines must intersect at the same point which is
the unique direction of the polarization of the incident spin. The
intersection point can also be seen in the spin space in Fig. \ref{fig3}(d).

Here the external magnetic field is fixed and can be integrated on the chip.
In fact, the three curves in the $(\theta, \varphi)$ plane always intersect
at the same point for any magnetic field. A proper magnetic field results in
better discrimination.

We note that the spin analyzer could be also achieved by a single SOC. In the
1D model, a single SOC only twists the spin in one direction ($x$ or $y$).
The incident angle can not be uniquely determined, there are always two
intersection points, no matter how many times we tune the strength of the SOC.
However, in the real 2D ring, the spin can be twisted more widely. The unique
intersection can appear. We show the numerical results in Fig. \ref{fig3}(e)
where only the Rashba SOC exists and the Dresselhaus SOC is absent. It is
clear that after three measurements with different strengthes of the Rashba
SOC, all the cones intersect at the unique intersection and the other
intersection has been lifted. So the spin polarization can also be identified
more easily.

\section{Conclusion}

In summary, we present a detailed study of the transport properties of the
device in which a quantum ring is in contact with two leads at the ends of
one diameter.  When the SOC is introduced, different phases are added in the
matter wave of the electron with different spins. So that the transmission
rates for different spins are no longer degenerate. By detailed analytical
and numerical studies, we find that in a simple quantum ring device, the spin
unpolarized current can be spin polarized parallel to the eigenstates of the
ring for appropriate SOC and the magnetic field. The direction of the
polarization can be tuned easily by the SOC and the magnetic field as well.
This simple device is therefore proposed to be a spin polarizer. Moreover,
similar to the light polarizer/analyzer, it can also be designed as an 
omnidirectional
all-electric spin analyzer by simply measuring the transmission rate of the
polarized incident via Eq. (\ref{tpj}). These findings pave the way to
control the system in spintronics and may be useful in quantum computation.
It also contributes an easy and controllable proposal to the design of
the high-performance all-electric transport device.

\section{Acknowledgement}

This work is supported by the NSF-China under Grant No. 11804396. J.S. is
supported by the NSF-China under Grant No. 11804397. A.G. acknowledges
financial support by the NSF-China under Grant No. 11874428, 11504066, and
the Innovation-Driven Project of Central South University (CSU) (Grant No.
2018CX044). F.O. acknowledges financial support by the NSF-China under Grant
No. 51272291, the Distinguished Young Scholar Foundation of Hunan Province
(Grant No. 2015JJ1020), and the CSU Research Fund for Sheng-hua scholars
(Grant No. 502033019). T.C. would like to thank Junsaku Nitta for helpful
discussion, in particular, for pointing out Ref.~\cite{Hatano}.

\appendix
\renewcommand\thefigure{S\arabic{figure}}
\renewcommand\theequation{S\arabic{equation}}
\setcounter{equation}{0}
\setcounter{figure}{0}

\section{Method and Formalism}

Here we explicitly derive the transmission rate in Eq.~(1).
We consider electrons as plane waves in the two leads with momentum $\hbar k$
and energy $E=\frac{\hbar^2k^2}{2m^{\ast}}$. Note that there is an
additional potential $U_0=-\beta_1^2$ induced by the Rashba SOC
\cite{sheng}, so that in the two arms, $E=\frac{\hbar^2k_0^2}{2m^{\ast
}}+U_0$. The wave vector in the two arms are \cite{AC3}
$k_{j}^{\mu}=k_{0}+j\left(  \frac{\Phi_{AB}}{2\pi r_{0}}+\frac{\Phi_{AC}^{\mu}%
}{2\pi r_{0}}\right).$
The incident current can be decomposed into the two eigenstates
$\chi^{1,2}$, and the electron is transported by this two channels.
We note that in this case the two channels are independent and there is no
interference between the two eigenstates. If the incident spin is polarized
along $\chi^\mu$ then the outcoming polarization is still along $\chi^\mu$.
If the incident current is decomposed into other two orthogonal states, then
the interference is difficult to deal with.

The wave function at the left lead contains the incident and the reflection,
$\Psi_{in}^{\mu}$. The wave function of the upper arm also contains two parts,
the clockwise and the anticlockwise movements, $\Psi_{1}^{\mu}$. In the same
manner, the wave function of the lower arm is $\Psi_{2}^{\mu}$. The output wave
function is marked $ \Psi_{out}^{\mu} $. All these wave functions are given by
\begin{eqnarray}
\Psi_{in}^{\mu} &=& \left(  e^{ikx}+r^{\mu}e^{-ikx}\right) \chi^{\mu}\left(
\pi\right), \\
\Psi_{1}^{\mu} &=& \sum_{j}C_{j}e^{jik_{j}^{\mu}x}\chi^{\mu}\left(
\phi\right), \\
\Psi_{2}^{\mu} &=& \sum_{j}D_{j}e^{jik_{-j}^{\mu}x}\chi^{\mu}\left( \phi\right), \\
\Psi_{out}^{\mu} &=& t_{\mu}e^{ikx}\chi^{\mu} \left(  0\right),
\end{eqnarray}
where $r^\mu$ is the reflection rate, $C,D$ are the parameters which can be
determined by the continuous condition, and $t_\mu$ is the variable
characterizing the transport properties of the device. The transmission rate
is thus given by $T_\mu= |t_\mu|^2$. By applying the Griffith boundary
conditions \cite{Griffith,AC1,Bellucci,Citro,Tang,Saeedia,Zhai},
the wave functions and the currents must be continuous at the two leads
($x=\pm r_0$ or $\theta=0,\pi$), we obtain six equations to solve the six
variables. Among them, the most wanted transmission rate can be solved,
\begin{equation}
T_{\mu}=\frac{K \Phi^\mu}{K K'+\left[4k_0^2\left(\Phi^\mu-K' \right)
+k^2\sin^2(\pi k_0r_0) \right]^2 },
\end{equation}
which is shown as Eq.~(1).

In order to calculate the transport properties numerically, it is convenient
to discretize the continuous Hamiltonian. We discretize $H$
on the sites of a square lattice with the lattice constant $a$ to obtain the
tight binding Hamiltonian. It is obtained by calculating the matrix elements
in the basis of position. The tight binding Hamiltonian is given by
\begin{align}
H  &  =\sum\limits_{i}^{}\left( V_{i}+4t+\frac{\Delta}{2}\sigma_{z}\right)
c_{i}^{\dagger}c_{i}-\sum_{\left\langle i,j\right\rangle }^{}\left(
t+s_{ij}\right) c_{i}^{\dagger}c_{j}^{}e^{i\theta_{ij}},  \label{tbhamiltonian}\\
t  &  =\frac{\hbar^{2}}{2m^{\ast}a^{2}},\\
s_{ij}  &  =-\frac{i\hbar g_{1}}{2a^{2}}\left( \sigma_{x}^{}\Delta
y-\sigma_{y}^{}\Delta x\right) -\frac{i\hbar g_{2}^{}}{2a^{2}}\left(
\sigma_{y}^{}\Delta y-\sigma_{x}^{}\Delta x\right) ,\\
\theta_{ij}  &  =\frac{e}{\hbar}\left( A_{x_{i}}\Delta x+A_{y_{i}}\Delta
y\right) ,
\end{align}
where $i$ runs over all sites, $\left\langle i,j\right\rangle $\,represents the
nearest neighbouring hopping only, $x_{i}$ and $y_{i}$ are the $x$ and $y$
coordinates of site $i$, and $\Delta x_{ij}=x_{j}-x_{i}$, $\Delta y_{ij}%
=y_{j}-y_{i}$. For convenience, we apply a hard-wall potential instead of the
parabolic potential,
\begin{equation}
V_{i}=%
\begin{cases}
0 & \left\vert r_{i}-r_{0}\right\vert \leqslant r_{w}\\
\infty & \left\vert r_{i}-r_{0}\right\vert >r_{w}%
\end{cases}
,
\end{equation}
where $r_{i}=\sqrt{x_{i}^{2}+y_{i}^{2}}$ and the width of the ring is $r_{w}$.
We connect two parallel leads to the ring, then the transmission properties
can be obtained by using the nonequilibrium Green's function (NEGF).

It is worthwhile to note that the continuous model and the tight binding model are
compatible and all the observable quantities in these two models are almost
equal (the small errors vanish when $a \rightarrow 0$). Moreover, the energy
spectrum has no essential difference in a
parabolic potential from that in a hard-wall potential, if $r_{w}$ matches the
confinement $\hbar\omega$. Then we consider  the transport properties in
such a lattice model with tight-binding Hamiltonian. The spin transmission
rate $T$ of the electron transporting from the left lead to the right lead
is defined by using the NEGF method \cite{dattabook,chang},
\begin{equation}
T_{\alpha}(E)=\text{Tr} \{\sigma_{\alpha}[ \Gamma
_{R}(E) G_{RL}(E) \Gamma_{L}(E) G_{LR}^{\dag}(E)]\},
%T_{\alpha}(E)=\text{Tr} \{\sigma_{\alpha}[ \Gamma(E) G(E) \Gamma(E) G^{\dag}(E)]\}
\end{equation}
where $ \alpha\in\left\{ x,y,z\right\} $, $\sigma_{x,y,z}$ are the Pauli
matrices and $\sigma_{0}$ is the unit matrix. The Green's function is defined
by the projection of the full Green's function \cite{dattabook},
\begin{eqnarray}
G_{RL} &=& P_R G P_L , \\
G (E) &=& (E - H -\Sigma_R -\Sigma_L )^{-1},
\end{eqnarray}
where $P_R, P_L$ are the projection operators to the right and the left
leads, $\Sigma_R,\Sigma_L$ are the self-energy of the right and the left
leads, respectively. The broadening function is defined by
$\Gamma_j = i \left[ \Sigma_j - \Sigma_j^\dag \right]$.

$T_{\alpha}(E)$ is the transmission rate of the $\left( \alpha\in
\left\{  x,y,z\right\}  \right) $ component of the spin or the total charge
transmission $\left( \alpha=0\right) $ while the energy of the incident
electron is $E$. Then the spin polarization rate $P$ is defined as:
\begin{equation}
P_{_{\alpha}}=\frac{T_{\alpha}}{T_0}\times100\%,\alpha
\in\left\{  x,y,z\right\}
\end{equation}
and $P_0=\sqrt{T_x ^{2}+T_y^{2}+T_z^{2} }/T_0=\sqrt{P_x^2+P_y^2+P_z^2}$.
$P_0$ represents the spin polarization of the outcoming electron. If $P_0=1$,
then the spin is fully polarized. If $P_0=0$, the spin is fully unpolarized.

%\section{Spin and current fields}

By diagonalizing the tight-binding Hamiltonian in Eq. (\ref{tbhamiltonian}),
we can have the value of the wave functions at each site, $\psi\left(
\mathbf{r}_i \right)$, which is a two component spinor. The physical
quantities can then be obtained. The spin fields are calculated by
\begin{equation}
\sigma_{\alpha}\left( \mathbf{r}_i\right) =\psi^{\dag}\left( \mathbf{r}_i\right)
\sigma_{\alpha}\psi \left( \mathbf{r}_i\right) ,\label{spin}
\end{equation}
and the density is given by $n\left( \mathbf{r}_i\right) =\psi^{\dag}
\left( \mathbf{r}_i\right) \psi \left( \mathbf{r}_i\right)$.
The average value of the observable quantity is thus given by $\left\langle
A\right\rangle =\sum_i \psi^{\dag} \left( \mathbf{r}_i\right) A
\psi^{\dag} \left( \mathbf{r}_i\right) \Delta x \Delta y$. The in-plane
field can be described by the vector field $\bm{\sigma}\left(
\mathbf{r}\right) =\left( \sigma_{x}\left( \mathbf{r}\right) ,\sigma_{y}\left(
\mathbf{r}\right) \right).$

The current operators can be derived by $j_{\mu}=-\frac{\delta H}{\delta A}$,
so that the on-site current densities are given by
\begin{align}
j_{x}\left( \mathbf{r}_i \right) & =\frac{e}{2m^{\ast}}\left[ \psi^{\dag}%
\left( \mathbf{r}_i\right) P_{x}\psi\left( \mathbf{r}_i\right) +\left( P_{x}
\psi \left( \mathbf{r}_i\right) \right) ^{\dag}\psi\left( \mathbf{r}_i\right)
\right] \nonumber\\
& -e\psi^{\dag} \left( \mathbf{r}_i\right) \left( g_{1}\sigma_{y}+g_{2}
\sigma_{x}\right) \psi\left( \mathbf{r}_i\right),\\
j_{y}\left( \mathbf{r}_i\right)  & =\frac{e}{2m^{\ast}}\left[ \psi^{\dag}%
\left( \mathbf{r}_i\right) P_{y}\psi\left( \mathbf{r}_i\right) +\left( P_{y}
\psi \left( \mathbf{r}_i\right) \right) ^{\dag}\psi\left( \mathbf{r}_i\right)
\right] \nonumber\\
& +e\psi^{\dag} \left( \mathbf{r}_i\right) \left( g_{1}\sigma_{x}+g_{2}
\sigma_{y}\right) \psi\left( \mathbf{r}_i\right).
\end{align}
The current is contributed by three parts,
\begin{equation}
j_{\alpha}\left( \mathbf{r}_i\right)  \equiv j_{z \uparrow, \alpha}\left(
\mathbf{r}\right) +j_{z \downarrow, \alpha}\left( \mathbf{r}\right)
+j_{SOC, \alpha}\left( \mathbf{r}\right),
\end{equation}
where
\begin{eqnarray}
j_{z \uparrow, \alpha}\left( \mathbf{r}_i\right)&=& \frac{e}{2m^{\ast}} \left[
\psi_{\uparrow}^* P_{\alpha}\psi_{\uparrow} +\left( P_{\alpha}\psi_{\uparrow}
\right) ^* \psi_{\uparrow} \right], \label{current+} \\
j_{z \downarrow, \alpha}\left( \mathbf{r}_i\right)&=& \frac{e}{2m^{\ast}} \left[
\psi_{\downarrow}^* P_{\alpha}\psi_{\downarrow} +\left( P_{\alpha}\psi_{\downarrow}
\right) ^* \psi_{\downarrow} \right], \label{current-}\\
j_{SOC, x} \left( \mathbf{r}_i\right)&=& -e\psi^{\dag}\left( g_{1}\sigma_{y}
+g_{2}\sigma_{x}\right) \psi, \label{currentx}\\
j_{SOC, y} \left( \mathbf{r}_i\right)&=& e\psi^{\dag}\left( g_{1}\sigma_{x}
+g_{2}\sigma_{y}\right) \psi. \label{currenty}
\end{eqnarray}
The on-site wave function spinor is
$\psi=
\left(
 \begin{array}{cc}
   \psi_{\uparrow} & \psi_{\downarrow} \\
 \end{array}
\right)^T$, and $\uparrow, \downarrow$ are related to the eigenstates of the
spin operator $\sigma_z$.

\section{Spin transmissions in different SOCs -- Spin filtering and spin
polarizer}\label{app2}

\begin{figure}[ptb]
\centering
\includegraphics[
height=3.7in,
width=3.2in
%scale=0.65
]%
{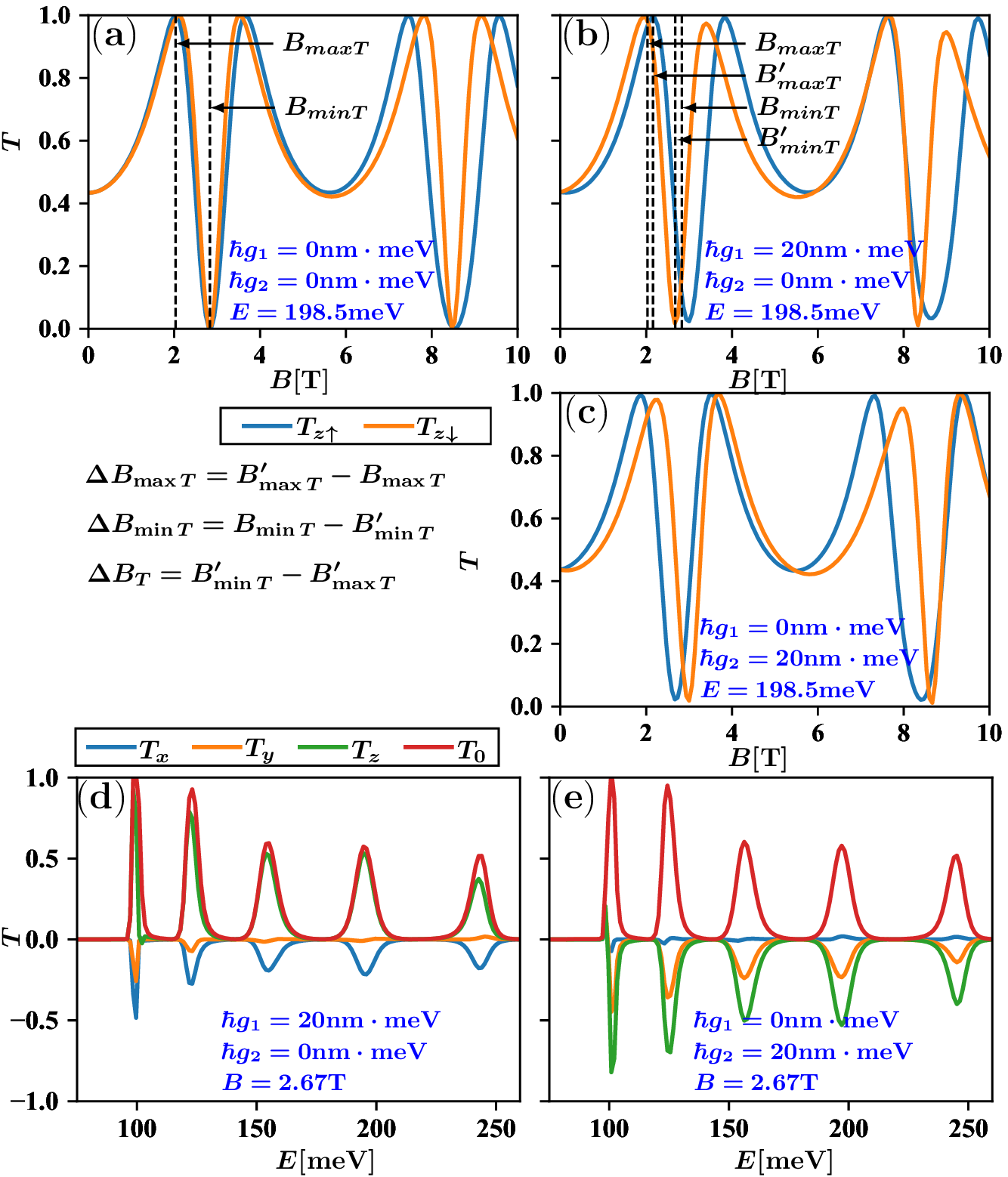}%
\caption{(color online). $T_{z\uparrow}$ and $T_{z\downarrow}$ (a) without SOC,
(b) with the Rashba SOC only, and (c) with the Dresselhauss SOC only.
$T_{\alpha}$ $\alpha\in\left( x,y,z\right)  $ (d) with the Rashba SOC only and
(e) with the Dresselhauss SOC only. }
\label{figapp1}
\end{figure}

We study in details how the transmission rate is related to the magnetic
field and the SOCs. We suppose that the input electrons are spin unpolarized.
If there is no SOC, the transported electrons are spin unpolarized as well,
i.e., $T_{z \uparrow}=T_{z \downarrow}$ for $g=0$. However, the Zeeman coupling
makes the transmission rate different, especially in a strong magnetic field,
as shown in Fig.~\ref{figapp1}(a). Since the minimum transmission rates for
spin up and down are all located at the same magnetic field, it would
be difficult to suppress one spin to zero and keep the other spin finite.

If the SOCs are introduced into the system, we find that the transmission
rate curves for spin down and spin up are well separated. If only the Rashba
SOC is present, the curve of $T_{z\downarrow}$ is shifted left and the curve
of $T_{z\uparrow}$ is shifted right as shown in Fig. \ref{figapp1}(b). If only
the Dresselhaus SOC exists, the shift of the curves is just opposite to that
in a Rashba ring, as shown in Fig. \ref{figapp1}(c).

If there is no magnetic field, the electron transports through the upper arm
and the lower arm with the same phase added ($T_1=T_2$), so that the
transmission rates for different spins depend on the ring itself and are
equal, as shown in Figs. \ref{figapp1}(b) and (c). However, in finite magnetic
fields the time-reversal symmetry is broken, the spin degeneracy in the ring
will be lifted more in the presence of either Rashba or Dresselhas SOC
strongly. Such a combination of the magnetic field and the SOCs can lead to
significant spin filtering effect.

%the SOC
%induces different phases adding for different spins. Hence, the transmission
%rates for different spins are separated.

In the numerical curves (Figs. \ref{figapp1}(b) and (c)), the spin filtering
appears periodically in a magnetic field, since the term $\Phi_{AB}+\Phi_{AC}$
in transmission rate Eq. (1) only depends on the magnetic field. For
instance, if only the Rashba SOC is present and the energy of the incident
electron is $E_{in}=198.5$ meV in Fig.~\ref{figapp1}(b), the lowest magnetic field
where the spin down is suppressed is at $B=2.67$T, which can be further lowered
by increasing the radius of the ring (at the same magnetic flux). In this
case, $T_{z\downarrow} \rightarrow 0$ and $T_{z\uparrow}$ is finite, so that
the output electrons are almost polarized to spin up, $P_z \rightarrow 1$.
For different energies of the input electrons, the transmission rates are
shown in Figs. \ref{figapp1}(d) and (e). The negative $T_{i}$ represents the $i$
component of the output spin is polarized in the negative direction of the
$i$ axis. In fact, according to Fig.~\ref{figapp1}(d) and the analysis of the 1D
model, the output spin is polarized along $\chi^{1,2}$ between the $z$ and $-x$
axis, since $T_y \approx 0$ and $T_{x,z}$ are finite.

%The Rashba SOC is able to tilt the $z$ component spin to the $x$ component
%spin, while the Dresselhaus SOC is able to flip the spin from the $z$ direction
%to the $y$ direction, as predicted in the 1D model analyze (see more
%discussions in Appendix \ref{app2}).

We now compare the transmission curves of the ring without SOC (Fig. \ref{figapp1}(a))
and the ring  with Rashba SOC (Fig. \ref{figapp1}(b)). The first maximum rate
for $T_{z\uparrow}$ is at $B_{maxT}$ in the ring without SOC. After the Rashba
SOC is set in, both of the $T_{z\uparrow}$ and $T_{z\downarrow}$ transmission
rates are shifted. We suppose that the first maximum value of $T_{z\uparrow}$
is shifted to $B'_{maxT}$. In the same manner, the first minimum rate in the
ring without SOC is at $B_{minT}=2.83$T, while the first minimum rate $T_{z
\downarrow}$ is shifted to $B'_{minT}$. We define the parameters
$\Delta B_{maxT}= B'_{maxT} -B_{maxT}$, $\Delta B_{minT}= B_{minT}-B'_{minT}$,
and $\Delta B_T= B'_{minT} -B'_{maxT}$ to study how the Rashba SOC shifts the
transmission rate curve and changes the polarization of the spin.

\begin{figure}[ptb]%
\centering
\includegraphics[
height=2.65in,
width=3.2in
%scale=0.6
]
{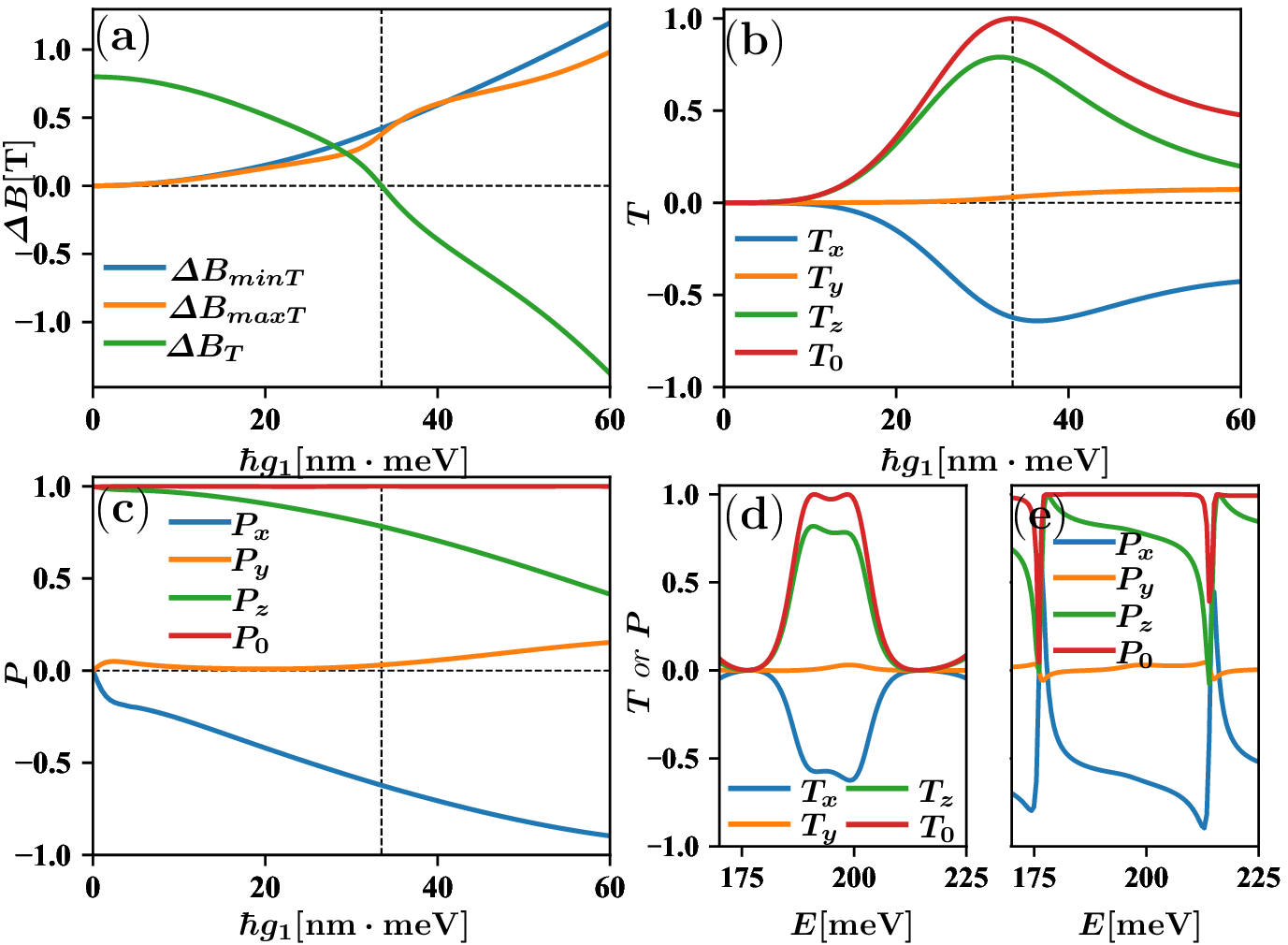}
\caption{(color online). (a) Magnetic field shifts $\Delta B$ induced by the
Rashba SOC. (b) Spin transmission rates $T$ and (c) spin polarization $P$ in
different $g_1$ with incident electron energy $198.5$ meV, in the magnetic
field where the minimum $T_{z\downarrow}$ is, so that $P_0=1$ always.
(d) The spin transmission rate
$T$ and (e) the spin polarization $P$ at $\hbar g_{1}$ = 33.5 nm$\cdot$meV and
$B$=2.41 T (corresponding to $\Delta B_T=0$) with energy.}%
\label{figapp2}%
\end{figure}

We show that in Fig. \ref{figapp2}(a) $\Delta B_T$ decreases with the increase
of the Rashba SOC $g_1$, due to the change of the AC phase $\Phi_{AC}^\mu$,
just as predicted in the 1D model.
%The shifts of the $T_{z\uparrow}$ and $T_{z\downarrow}$ are more or less
%equal, $\Delta B_{minT} \approx \Delta B_{maxT}$, since the AB phase is weak.
When $\hbar g_1=33.5$ nm$\cdot$ meV, $\Delta B_T =0$, which means that the
transmission rate of spin down is suppressed to minimum and the transmission
rate of spin up is maximum. Interestingly, at this point the total charge
transmission rate is exactly $1$ as shown in Fig. \ref{figapp2}(b). Hence,
$\chi^1$ is completely suppressed, and $\chi^2$ passes the ring freely.
Meanwhile, the $y$ component of the spin is almost suppressed, $T_y
\approx 0$, and both $|T_x|$ and $|P_x|$ increase, as shown in Figs.
\ref{figapp2}(b) and (c). We are then able to control the direction of the spin
polarization by tuning the Rashba SOC. The tunable spin polarizer is thereby
established. In Fig.~\ref{figapp2}(d), we show how the spin polarization and the
transmission rate vary with the energy of the incident electron. Basically,
$T_y$ is close to 0 and $P_y$ is also always very small. It is nonzero
comparing with the 1D model, due to the Zeeman coupling and the width of the
ring. In Figs.~\ref{figapp2}(d) and (e) we show that if the energies of the
incident electrons are in the region $[185,205]$ meV, the spin transmission
and polarization are stable. So that the outcoming current which is obtained
by integral of the transmission rate $T$ over this region is almost fully
spin polarized. In order to exclude the unwanted transmission below
$185$meV, we can apply a gate to lift the whole energy band of the lead.

In our ring device, the Rashba SOC tilts the spin to the $x$ axis, while the
Dresselhaus SOC flip the spin towards the $y$ direction. It can also be
understood simply as follows. When the magnetic field is absent, the effective
vector potential induced by the SOCs is
\begin{eqnarray}
  A_x^{SOC} &=&-\frac{m}{e \hbar} (g_1 \sigma_y + g_2 \sigma_x), \\
  A_y^{SOC} &=& \frac{m}{e \hbar} (g_1 \sigma_x + g_2 \sigma_y).
\end{eqnarray}
Suppose the incident wave function is spin polarized, $\psi^{in}_+=
(\begin{array}{cc}
1 & 0
\end{array})^T$, then the outcoming wave function influenced by the SOC is
given by $\psi^{out} \propto e^{-i A_x \cdot 2(r_0+r_w)} \psi^{in}$, since
the coordinate difference in the $y$ direction is zero. If there is only
Rashba existing,
\begin{equation} \label{psiout}
  \psi^{out}_R\propto (1+ i \gamma g_1 \sigma_y  )
  \left(
     \begin{array}{c}
       1 \\
       0 \\
     \end{array}
  \right)=\left(
     \begin{array}{c}
       1 \\
       -\gamma \\
     \end{array}
  \right),
\end{equation}
where $\gamma = 2(r_0+r_w)\frac{m}{e \hbar} g_1 > 0$. So we have $\langle
\sigma_x \rangle= -2\gamma < 0$ and $\langle \sigma_y \rangle=0$. The spin is
torqued from the $+z$ direction to $-x$. If the incident electron is spin
down, $\psi^{in}_-=(\begin{array}{cc}
                       0 & 1
                     \end{array})^T
$, then $\psi^{out}_R=(\begin{array}{cc}
                   \gamma & 1
                 \end{array})^T
$, and then $\langle \sigma_x \rangle= 2\gamma > 0$ and $\langle \sigma_y
\rangle=0$. In Fig. \ref{figapp2}(d), however spin down is suppressed in the
transport, and the spin up $\psi^{in}_+$ is flipped to the $-x$ axis. On
the other hand, if only the Dresselhaus SOC is present, we can do the same
calculation.  For $\psi^{in}_+$, $\psi^{out}_D=(\begin{array}{cc}
                               1 & i \gamma
                             \end{array})^T
$, so that $\langle \sigma_x \rangle= 0$ and $\langle \sigma_y \rangle=2
\gamma>0$. For $\psi^{in}_-$, $\psi^{out}_D=(\begin{array}{cc}
                               i \gamma & 1
                             \end{array})^T
$, so that $\langle \sigma_x \rangle= 0$ and $\langle \sigma_y \rangle=-2
\gamma<0$. In Fig. \ref{figapp2}(e), the spin up is suppressed, while the spin
down $\psi^{in}_-$ is flipped to the $-y$ axis in the transport. The analysis
agrees with the numerical results perfectly.

\begin{figure*}[ptb]%
\centering
\includegraphics[
scale=0.85]%
{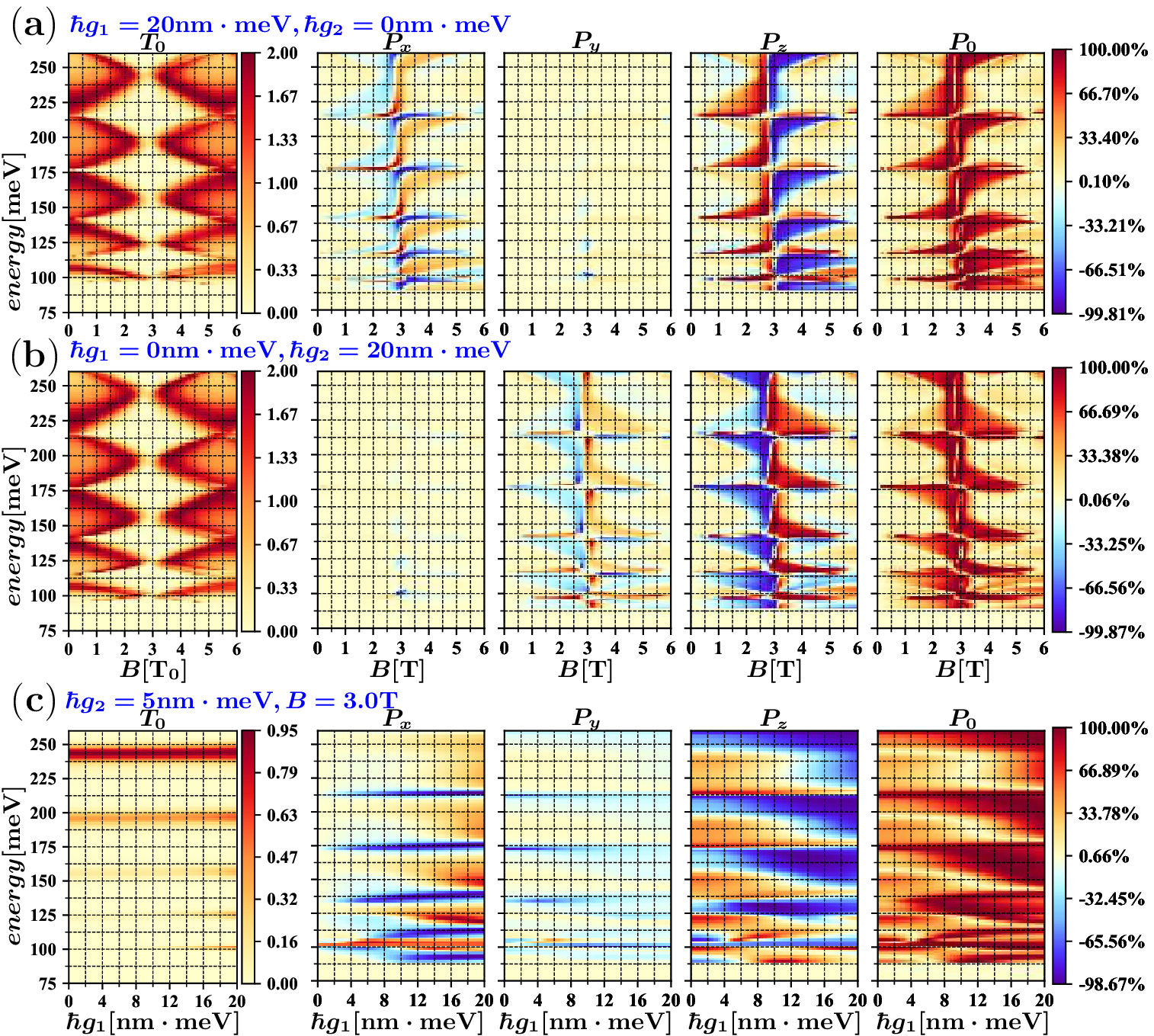}%
\caption{(color online). The\ transmission $T$ and spin polarizability
$P$  with (a) Rashba SOC only, (b) Dresselhauss SOC only, and (d) fixed
Dresselhauss SOC but tunable Rashba SOC.}
\label{figapp3}
\end{figure*}

As drawn in Fig. \ref{figapp3}, we show the relation between the spin
transmission rates and the spin polarizations for different SOCs. If only
the Rashba SOC is existing, then $T_y$ and $P_y$ will be suppressed shown in
Fig. \ref{figapp3}(a). The direction of the spin polarizer can be tuned by the
strength of the Rashba SOC in the plane $xOz$. If only the Dresselhaus SOC is
present, then $T_x$ and $P_x$ will be suppressed
shown in Fig. \ref{figapp3}(b). The direction of the spin polarizer is then in
the plane $yOz$. If both of the SOCs are present, then the situation becomes
complicated and spin polarizer can be controled more widely, as shown in Fig.
\ref{figapp3}(c). However, we find that if the outcoming spin needs to be
polarized well, then it is better to keep one SOC dominating the system. The
competition of the two SOCs makes the spin more difficult to be polarized.

\section{Spin textures and current in the transport} \label{app3}

\begin{figure*}[ptb]
\centering
\includegraphics[
scale=0.8]%
{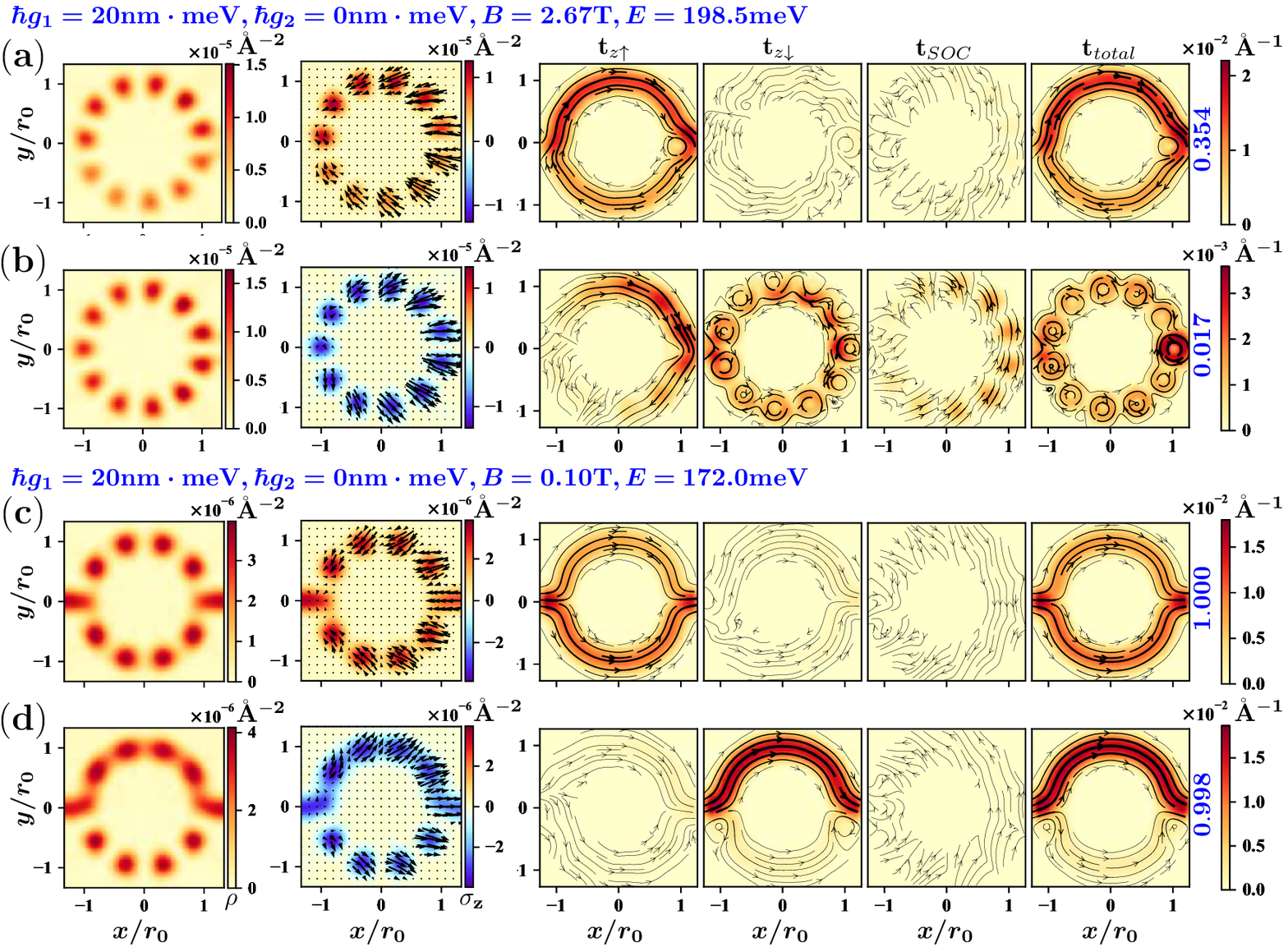}%
\caption{(color online). Transport status and transmission flow density for
electrons with fixed incident energy and magnetic fields in a Rashba ring.
The first two columns (from left to right) show the charge and spin fields
when the transport reaches the equilibrium status, the colors represent the
density and $\sigma_z(\mathbf{r})$, respectively. In (a) and (b), strong
spin filtering is found, i.e. the transmission rate of spin up is much higher
than that of spin down. In (c) and (d) the outcoming spin is fully
unpolarized although the total transmission rate is almost 1. For (a) and (b),
$\hbar g_1=20$ nm$\cdot$ meV, $B=2.67$ T and the incident energy is $198.5$ meV.
For (c) and (d), $\hbar g_1=20$ nm$\cdot$ meV, $B=0.1$ T and the incident energy
is $172$ meV. }
\label{figapp4}%
\end{figure*}
%EndExpansion

The incident spin is supposed to be unpolarized, so that the wave function of
the incident electrons $\psi^{in}$ can be decomposed to two parts in any
direction of the spin polarization. Without the loss of generality, we decompose
the incident electron in the basis of $\sigma_z$, $|\psi^{in}_{z\uparrow}|^2=
|\psi^{in}_{z\downarrow}|^2$. The spins of the two parts are independently
polarized along $z$ or $-z$ direction, respectively. For each part of the
incident electron, it contributes one transmission channel in the transport.
Then we can figure out which channel plays more important role in the
transport. The wave function of the incident electron is supposed to be the
wave function of the lowest band of the lead. By employing Eq. (7) we can 
obtain the wave function in the ring by the Green's function method,
\begin{equation}
\psi^{ring}= G \tau_L \psi^{in}_{z\uparrow(\downarrow)},
\end{equation}
where $\tau_L$ is the coupling matrix between the incident (left) lead and
the ring \cite{dattabook}. We again employ the current densities $\mathbf{j}
_{z\uparrow},\mathbf{j}_{z\downarrow}$ and $\mathbf{j}_{SOC}$ defined in Eqs.
(\ref{current+}) to (\ref{currenty}), where $\psi$ needs to be replaced by
$\psi^{ring}$. We can define the transmission density $t_\alpha (\mathbf{r})$
proportional to the current,
\begin{equation}\label{tdensity}
t_\alpha (\mathbf{r}) = - \frac{\hbar}{2e \cdot \text{meV}} \mathbf{j}_\alpha
(\mathbf{r}).
\end{equation}
The transmission rate is thus obtained by $T_\alpha = \sum_{i} t_\alpha
(\mathbf{r}_i)$, where $i$ includes all the sites between the lead and the
ring.

For simplicity, we consider  only the Rashba SOC in two different cases: (i)
$\hbar g_1=20$ nm$\cdot$ meV at $B=2.76$T, the electron of the incident energy
$E_{in}=198.5$ meV has the transmission rate $T_0=0.372$; (ii) $\hbar g_1=20$
 nm$\cdot$ meV at $B=0.1$T, the transmission rate of the electron with $E_{in}
=172$ meV is $T_0=1.998$. In Figs. \ref{figapp4}(a) and (b), we show how the
incident wave functions $\psi^{in}_{z \uparrow}=(\begin{array}{cc}
                                     1 & 0
                                   \end{array}
)^T$ and $\psi^{in}_{z \downarrow}=(\begin{array}{cc}
                                     0 & 1
                                   \end{array}
)^T$ are transported through the ring, respectively, where the outcoming
spin is polarized and the transport rate is relatively low. In Figs.
\ref{figapp4}(c) and (d), we show how the incident wave functions transport
in the ring when the magnetic field is $B=0.1$T, where the electrons pass
through the ring freely but the spin is not polarized at all.

When the transport reaches the equilibrium status,
the spin and charge densities and the current densities are plotted in Fig.
\ref{figapp4}. Both the charge densities and the spin textures shown in the
first two columns (from left to right) of Fig. \ref{figapp4} are periodically
distributed in the ring as a stationary wave, due to the interference of the
matter wave of the electron. The spin textures also support the analysis of
the outcoming spins derived in Eq. (\ref{psiout}), i.e. at the right lead
$\langle \sigma_y \rangle=0$, the $x$ component spin is generated in the
transport by the SOC and the direction of $\sigma_x (r)$ depends on the
polarization of the incident spin.

Comparing with the case without the magnetic field \cite{sun,datta}, here the
vector potential of the external magnetic field and the effective vector
potential induced by the SOC give different phases to the upper and the lower
arms, respectively. This phase difference leads to different transmission for
different spins and can be observed by the transport experiment.

In the spin up channel in case (i), electron is mostly transported by the
current $\mathbf{j}_{z\uparrow}$, which means the SOC does not contribute a
lot in the transmission. In the spin down channel, the SOC flips spin and
induces stronger transmission. However, the transmission of this channel is
still weak, only contributes $1/20$ of the spin up channel. In this case, the
spin is thus strongly polarized. In the case (ii), both of the two channels
have high transmission rate, close to $1$. The current in the spin down
channel is obviously imbalanced in the upper and lower arms. However the
outcoming spin has half in spin up and half in spin down, which means that
the ring in this case is good in transport but fails to polarize the spin.

There are circular currents in the ring, which do not contribute to total
transmission, when the transmission rate is low. It keeps the current
conserved. If the transmission is high, the internal circling is weak, but
the imbalance between the currents of the upper and the lower arms is
explicit. From the detailed transport pictures shown in Fig. \ref{figapp4},
we can clearly see how the electron passes through the ring. This method is
general and can be applied to other systems as well.


\begin{thebibliography}{99}
\bibitem{latticeconstant} The lattice constant of InAs is about
$0.6$nm, however, the results of $1$nm lattice are no significantly different
from the real physical system. For simplicity and saving computing resources
we utlized a larger lattice constant.
\end{thebibliography}
\end{document}